\documentclass[twocolumn,floatfix,prl,showpacs]{revtex4}
\usepackage{graphicx}
\usepackage{amssymb}
\usepackage{amsmath}
\usepackage{color}
\usepackage{psfrag}
\usepackage{epsfig}
\usepackage{bbm}
\usepackage{bm}
\newcommand{\ket}[1]{| #1 \rangle}

\newcommand{\rb}[1]{\left( #1 \right)}

\newcommand{\ew}[1]{\langle #1 \rangle}
\newcommand{\beq}{\begin{eqnarray}}
\newcommand{\eeq}{\end{eqnarray}}

\newcommand{\op}[2]{| #1 \rangle \langle #2 |}

\newcommand{\eq}[1]{Eq.~(\ref{#1})}
\newcommand{\fig}[1]{Fig.~\ref{#1}}

\newcommand{\bs}[1]{\boldsymbol{#1}}

\begin{document}
\title{Measuring the Entanglement between Double Quantum Dot Charge Qubits}
\author{Clive Emary}
\affiliation{
  Institut f\"ur Theoretische Physik,
  Hardenbergstr. 36,
  TU Berlin,
  D-10623 Berlin,
  Germany
}

\date{\today}
\begin{abstract}
  We present a scheme for creating and measuring entanglement between two double quantum dot charge qubits in a transport set-up in which voltage pulses can modify system parameters.  Detection of entanglement is performed via the construction of a Bell inequality with current correlation measurements.  An essential feature  is the use of the internal dynamics of the qubits as the constituent electrons tunnel into the leads to give the single-particle rotations necessary for the Bell measurement.
\end{abstract}
\maketitle

In an important recent experiment \cite{shi09}, Shinkai {\it et al.}  have demonstrated correlated coherent oscillations between two coupled double quantum dot (DQD) charge qubits formed in a top-gated semiconductor heterostructure \cite{shi07}.  From transport measurements, indications were obtained that it should be possible to perform a suite of universal two-qubit quantum gates with such a setup.  However, such a claim can only be substantiated if it can be shown that the operations can entangle the qubits \cite{gates}. This brings us to the question that is the focus of this Letter: If the qubits of Shinkai et al were entangled, how could we tell? Is it possible to detect and measure the entanglement between DQD charge qubits in a transport set-up such as that of Ref.~\cite{shi09}?

We answer these questions here by describing a series of shotnoise measurements that can be used to construct a Bell's inequality (BI) \cite{bell64}, the violation of which provides a clear signal of, and in certain circumstances quantitative information about, the entanglement between the qubits.  
Shotnoise and the Bell's inequality have been combined to study entanglement in mesoscopic systems before \cite{mesoBell}. However, the system here differs in several respects.  Most important is that here we are in the sequential-tunnelling regime and coherence between the electrons is assumed to be lost once the electrons tunnel to the leads. 
Furthermore, the BI requires single-qubit rotations and it is not initially obvious how this may be accomplished.
In this Letter, we show how these obstacles can be overcome by making use of the internal dynamics of the qubits.  As the qubit electrons tunnel into the leads, i.e. as the
qubits decay, they experience the action of the system Hamiltonian and this rotates the qubits.  We show how an appropriate set of current correlation measurements can extract the relevant information from the stochastic background of the qubit decay.

\begin{figure}[tb]
  \begin{center}
  \epsfig{file=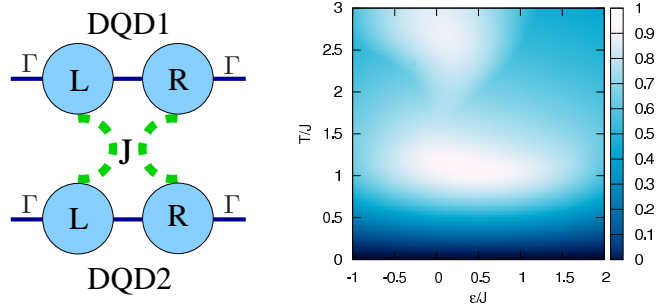, clip=true,width=\linewidth}
  \caption{
      (a) The system consists of two DQDs with each dot connected to its respective lead.  With one electron in each DQD, two charge qubits are formed that interact with interaction strength $J$.
      (b)  The maximum concurrence ${\cal C}$ of the pure two-qubit state evolving from state $\ket{LL}$ at $t=0$ under the action of $H_\mathrm{DQD}$. Only the time interval $[0, \tau_\mathrm{max}]$ with $ \tau_\mathrm{max} = J^{-1}$ is considered.  Other parameters were $\epsilon^{(2)}=\epsilon^{(1)}=\epsilon$ and $T^{(2)}=T^{(1)}=T$.
       \label{FIG1}
 }
  \end{center}
\end{figure}

The set-up of interest is sketched in \fig{FIG1}a with four quantum dots (QDs) making up two charge qubits, with
the position of the electron within a DQD (left or right) corresponding to the two logical qubit states.
With a pseudo-spin convention that $\ket{\downarrow}\equiv\ket{L}$ and $\ket{\uparrow}\equiv\ket{R}$, the Hamiltonian of the isolated two-qubit system is 
\beq
  H_\mathrm{DQD} 
  &=&
  \sum_{i=1,2} 
  \rb{
    \epsilon_i \sigma_z^{(i)} + T_i \sigma_x^{(i)}
  }
  +J \sigma_z^{(1)}\otimes \sigma_z^{(2)}
  \label{DQDHam}
  ,
\eeq
with $\epsilon_i$ and $T_i$ the detuning and tunnel coupling of DQD $i=1,2$, and with $J$ the strength of the Coulomb interaction between the electrons.

We assume that the DQD parameters as well as the chemical potentials of the leads can be controlled to a limited extent via voltage pulses and, as in Refs.~\cite{hay03,shi09}, we envisage an experiment that consists of a series of repeated steps.  At the start of the cycle, both DQDs are empty.
Then the left chemical potentials are raised, such that one electron tunnels into each of the dots from the left.
As in Refs.~\cite{hay03,shi09}, maintaining a large bias across the dots produces detunings $\epsilon_i$ such that the electrons tunnel into localised states $\ket{L^{(i)}}$. 
The right chemical potentials are then raised such that the electrons are trapped and two qubits are formed. They remain trapped in the system with the full two-qubit Hamiltonian acting on them for a time  $\tau_\mathrm{init}$, at the end of which the two-qubit system is left in the state $\rho_0$.  We want to measure the entanglement of this state.  Finally, all chemical potentials are dropped below the dot levels and the electrons escape into the leads. Sufficient time is waited for this to have happened with certainty before the sequence is repeated.

The only control over dot parameters assumed here is that they be set once during the initialisation of state $\rho_0$, and subsequently to a different set for the detection phase.  It is essential that various different configurations are possible so that there is a range of single-particle rotations for the BI measurement.
As we show below, with an appropriate choice of current correlation measurements, we can construct the correlator ${\cal E}(\bs{a},\bs{b}) \equiv  \ew{ \bs{a}\cdot \bs{\sigma}^{(1)} \bs{b}\cdot \bs{\sigma}^{(2)}}$ with unit vectors $ \bs{a}$ and $ \bs{b}$ determined by the DQD parameters.  Repeating the experiment with different parameters allows us to build the correlator
\beq
  F \equiv
 {\cal E}(\bs{a}, \bs{b})
  + {\cal E}(\bs{a},\bs{b'})
  + {\cal E}(\bs{a'},\bs{b})
  - {\cal E}(\bs{a'},\bs{b'})
  .
\eeq
The pertinent BI is the CHSH inequality, which reads $ - 2 \le F \le 2$ \cite{CHSH69}.  A measurement of  $|F|>2$ therefore indicates entanglement.  Maximising $|F|$ with respect to vectors $\bs{a}, \bs{b},\bs{a'},\bs{b'}$ yields  $|F_\mathrm{max}|$, and for pure states we have $|F_\mathrm{max}|=2 \sqrt{1+{\cal C}^2}$ \cite{pop92}, with ${\cal C}$ the concurrence, a measure of two-qubit entanglement \cite{woo98}.  
For mixed states, the region of violation is bounded  by
$
	\mathrm{max}\left[ 2,2\sqrt{2}{\cal C} \right] 
	\le
	|F_\mathrm{max} |
	\le 
	2 \sqrt{1+{\cal C}^2}
$ \cite{hor95,ver02}.

{\em Initialisation.}  We first discuss the nature of the entangled states that can be produced within this scheme before analysing the detection protocol in more detail.  
The Hamiltonian of \eq{DQDHam} can be used to form many different quantum gates.  For example, with $\epsilon_1 = - \epsilon_2$, $T_1 = T_2$, and $|J \pm \epsilon_1|\gg T_1$,  the Hamiltonian is effectively that of a FLIP gate,
$
  H_\mathrm{DQD} \approx H_\mathrm{FLIP} 
  = \textstyle{\frac{1}{2}}\Omega \rb{\op{LL}{RR}+\op{RR}{LL}}
$,
with effective coupling
$\Omega = 2T_1^2 J / (J^2 -\epsilon_1^2)$ \cite{shi09}.
Starting with the system in state $\ket{LL}$, and allowing it to evolve for a time $\tau$, the density matrix of the system becomes
\beq
  \rho(\tau) &=& \frac{1}{2}
  \left\{
    (1+\alpha) \op{LL}{LL} + (1-\alpha) \op{RR}{RR}
  \right.
    \nonumber\\
    &&
  \left.
    + i {\cal C} (\op{LL}{RR} - \op{RR}{LL})
  \right\}
  \label{rhoFLIP}
  ,
\eeq
with $\alpha = \cos (\Omega \tau)$ and ${\cal C} =  \sin(\Omega \tau) e^{-\gamma \tau}$, the concurrence of the state, in which we have included a dephasing at rate $\gamma$.
These states, although mixed for finite $\gamma$, have the same violation of the Bell inequality as a pure state of the same concurrence, $F(\rho) = 2 \sqrt{1+{\cal C}^2}$ \cite{ver02}.
The entangled states obtained by letting $H_\mathrm{DQD}$ act on the state $\ket{LL}$ for a time are not restricted to these FLIP-class states.  The entanglement of this more general class of states is investigated in \fig{FIG1}b.  Clearly, a high degree of entanglement (concurrence approaching unity) is obtainable under reasonable conditions.

{\em Entanglement detection}.
Let us initially consider an ideal model of the decay of the two-qubit system.  We assume that, once the initial entangled state is created, the interaction between the qubits is turned off ($J=0$). 
This will be a good approximation if the coupling of the QDs to the leads is strong enough that the `measurement phase' is quick compared to the interaction timescale $J^{-1}$.
Let us also assume the most general form for the single-qubit Hamiltonian
$H^{(i)} = \Lambda^{(i)} \bs{n}^{(i)}\cdot\bs{\sigma}^{(i)}$,
where $\Lambda^{(i)}$ is the single-qubit energy scale, $\bs{n}^{(i)}$ is a unit vector, and $\bs{\sigma}^{(i)}$ is a vector of Pauli matrices.
We return to our more realistic DQD model shortly.

We consider the DQD system to be in the strong Coulomb blockade regime, such that at most one excess electron is permitted in each DQD.
We describe the tunnelling of electrons with a Markovian master equation, $\dot{\rho} = {\cal L}\rho$, with $\rho$ the density matrix of the qubit pair and ${\cal L}$ the total system Liouvillian.  The initial state of the system is $\rho_0$, some entangled state such as that of \eq{rhoFLIP}.
In the noninteracting approximation ($J=0$), we can analyse the dynamics of each qubit separately and need only consider the two qubits together when we introduce the initial state.
To facilitate calculation of current statistics, we introduce counting fields $\chi_L^{(i)}$ and $\chi_R^{(i)}$ of electrons in the $L$ and $R$ leads of dot $i=1,2$ \cite{fcs} to obtain the $\chi$-resolved master equation: $\dot{\rho}^{(i)}(\chi) = {\cal L}^{(i)}(\chi)\rho^{(i)}(\chi)$ with 
${\cal L}^{(i)}(\chi) = {\cal L}^{(i)}(\chi_L^{(i)},\chi_R^{(i)})$, the  $\chi$-resolved Liouvillian of dot $i$, and similarly for the density matrix, 
which includes the empty state $\ket{0^{(i)}}$ as well as the two qubit states $\ket{L^{(i)}}$ and $\ket{R^{(i)}}$. The Liouvillian can be written
${\cal L}^{(i)}(\chi) = L^{(i)}_0 + \Sigma^{(i)}(\chi)$ with first term describing the internal dynamics
$
  L^{(i)}_0 \rho^{(i)} =  -i \left[ H^{(i)},\rho^{(i)} \right]
$,
and the second, the coupling to the leads. 
With chemical potentials set far below the dot levels, the latter can be written
\beq
   \Sigma^{(i)} (\chi)\rho^{(i)}
    &=&
    -\frac{1}{2} \sum_{\alpha=L,R} \Gamma_{\alpha}^{(i)} 
    \left\{
      {s_\alpha^{(i)}}^\dag   s_\alpha^{(i)}\rho^{(i)}
    \right.
    \nonumber\\
    &&
    ~
    \left.  
      +  \rho^{(i)} {s_\alpha^{(i)}}^\dag s_\alpha^{(i)}
      -2 s^{(i)}_\alpha \rho^{(i)} {s_\alpha^{(i)}}^\dag e^{i\chi_\alpha}
    \right\}
    ,
    \label{kernelCHI}
\eeq
with operators $s_\alpha=\op{0}{\alpha^{(i)}}$ describing the jump on an electron from localised state $\alpha^{(i)}=L^{(i)},R^{(i)}$ to the leads.  Setting $\chi_\alpha=0$ in \eq{kernelCHI} we obtain the familiar master equation in Lindblad form which describes the evolution of the actual density matrix $\rho^{(i)}$.
From this point on, we assume that all rates are identical, $\Gamma^{(i)}_L=\Gamma^{(i)}_R=\Gamma$, and assume that this rate is faster than dephasing rate $\gamma$, such that such dephasing from external sources can be neglected in the detection phase (see later).

In Laplace space, the density matrix of the system at arbitrary time is
$
  \rho^{(i)}(\chi;z) =  \Omega^{(i)}(\chi;z) \rho_0
$ with the $\chi$-resolved propagator for a single DQD,
$
  \Omega^{(i)}(\chi;z) = \left[z-L_0^{(i)} -\Sigma^{(i)}(\chi)\right]^{-1}
$.
We are only interested in the situation in which the system starts with one electron in each DQD, in which case, the propagator in the long time limit reduces to
\beq
  \Omega_\infty^{(i)} (\chi) = 
  \frac{{\cal J}^{(i)}_L e^{i \chi_L^{(i)}} +   {\cal J}^{(i)}_R  e^{i \chi_R^{(i)}}  }{2}
  \rb{
    \mathbbm{1}^{(i)} + {\cal R}(\theta^{(i)}, \bs{n}^{(i)})
  }
  \label{propLTL}
  .
\eeq
Here, ${\cal J}_\alpha^{(i)}$ are jump super-operators defined by ${\cal J}_\alpha^{(i)} \rho^{(i)} = \op{0^{(i)}}{\alpha^{(i)}}\rho^{(i)}\op{\alpha^{(i)}}{0^{(i)}}$ and $ {\cal R}(\theta^{(i)}, \bs{n}^{(i)}) \rho^{(i)} = U^{(i)} \rho^{(i)}{U^{(i)}}^\dag$ with
$
  U^{(i)} = \exp\rb{-\frac{i}{2} \theta^{(i)} \bs{n}^{(i)}\cdot \bs{\sigma}^{(i)}}
$
a unitary rotation about axis $\bs{n}^{(i)}$ by angle 
$
  \theta^{(i)} = 
  \textstyle{\frac{1}{2}} \arctan \rb{2 \Lambda^{(i)} / \Gamma}
$.
The above form of the propagator is the main formal result of this work.  It shows that, in the long-time limit, the behaviour of the system effectively decomposes into two parts: one in which the qubits decay directly, and one in which the qubits are first rotated and then leave the dots.  
This rotation originates from the action of the single-qubit Hamiltonian $H^{(i)}$ acting for a time governed by the ratio of $\Lambda^{(i)}$ to $\Gamma$.

The moment generating function (MGF) for the two-qubit system in the long-time limit is
$
  {\cal M}(\chi) = \mathrm{Tr}\left\{\Omega_\infty^{(1)}(\chi) \otimes\Omega_\infty^{(2)}(\chi) \rho_0\right\}
$,
where $\rho_0$ is the two-qubit entangled state.  The first moment,
${\cal M}^{i}_X \equiv \left.\partial {\cal M}/\partial (i\chi_X^{(i)})\right|_{\chi\to 0}$, corresponds to the mean number of electrons transferred to lead $X=L,R$ of dot $i$ in the measurement part of the cycle. More importantly, let us define
$
  {\cal M}_{XY} \equiv
  \left.
    \partial^2 {\cal M}/\partial(i\chi^{(1)}_X)\partial(i\chi^{(2)}_Y)
  \right|_{\chi \to 0}
$
as the cross correlator between the number of electrons emitted into lead $X$ of dot $1$ and into lead $Y$ of dot $2$.  This quantity can be extracted from shotnoise measurements \cite{mesoBell} as the preparation-detection cycle is repeated continuously.  
In analogy to the standard CHSH measurement \cite{CHSH69}, let us define
\beq
  C({\cal R}^{(1)},{\cal R}^{(2)}) 
  &\equiv&
  {\cal M}_{LL} 
  -{\cal M}_{LR} 
  -{\cal M}_{RL}  
  +{\cal M}_{RR} 
  \nonumber\\
  &=&\frac{1}{4}
  \mathrm{Tr}\left\{
  \sigma_z^{(1)} \otimes \sigma_z^{(2)} 
  \rb{
	\mathbbm{1}^{(1)} + {\cal R}(\theta^{(1)}, \bs{n}^{(1)})
  }
  \right.
  \nonumber\\
  &&
  \left.
  \times
  \rb{
	\mathbbm{1}^{(2)} + {\cal R}(\theta^{(2)}, \bs{n}^{(2)})
  }
  \rho_0
  \right\}
  \label{Cfn}
  .
\eeq
This quantity still has contributions from the non-rotating decay. However, if we take the following combination of $C$-correlators:
\beq
  {\cal E} (\bs{a},\bs{b}) 
  &\equiv&
  4C({\cal R}^{(1)},{\cal R}^{(2)}) 
  -2C({\cal R}^{(1)},\mathbbm{1}^{(2)})
  \nonumber\\
  && -2C(\mathbbm{1}^{(1)},{\cal R}^{(2)}) 
  +C(\mathbbm{1}^{(1)},\mathbbm{1}^{(2)}) 
  ,
\eeq
we obtain
$
  {\cal E} (\bs{a},\bs{b}) 
  =  \ew{
     \bs{a}\cdot \bs{\sigma}^{(1)} \bs{b}\cdot \bs{\sigma}^{(2)}
  }
$
with $\bs{a}\cdot \bs{\sigma}^{(1)}  =  {U^{(1)}}^\dag\sigma_z^{(1)}U^{(1)}$ and similarly for $\bs{b}$.  This series of measurements then yields exactly the correlation function required to form the CHSH inequality.  The operations ${\cal R}^{(i)}=\mathbbm{1}^{(i)}$ can be realised simply by any rotation about the $z$-axis, i.e. with $\epsilon_i \gg T_i$.

\begin{figure}[tb]
  \begin{center} 
  \epsfig{file=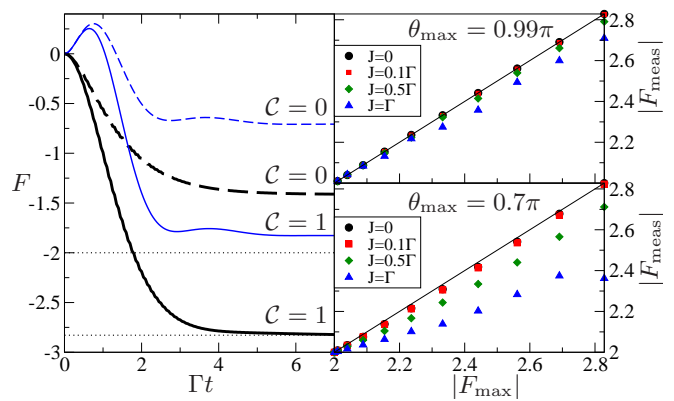, clip=true,width=\linewidth}
    \caption{
    {\bf (a)}
    Time evolution of CHSH correlator $F$ for two initial FLIP-class states (\eq{rhoFLIP}): one separable (${\cal C}=0$, dashed lines) and one maximally entangled (${\cal C}=1$, continuous  lines) with $J=0$.  The thick lines show $F$ obtained with vectors close to those yielding maximum CHSH violation for ${\cal C}=1$.
    The thin lines show $F$ with a choice of vectors such that the inequality is never violated.
    The violation (or otherwise) becomes apparent after a time $\sim 4\Gamma^{-1}$.
    {\bf (b)},{\bf (c)} Plot of $|F_\mathrm{meas}|$ vs $|F_\mathrm{max}|$ for FLIP-class states, with  $|F_\mathrm{meas}|$ obtained the under constraint $0 \le \theta \le \theta_\mathrm{max}$, and with finite interaction strength $J$.
    The straight line indicates $|F_\mathrm{meas}| = |F_\mathrm{max}|$
    \label{FIG2}
  }
  \end{center}
\end{figure}

Since we start with exactly one electron in each DQD, in the long time limit the total number of electrons to leave each DQD must also be unity.  This means that we can re-express the correlation function of \eq{Cfn} as
$
  C({\cal R}^{(1)},{\cal R}^{(2)}) 
  =
  1 + 4 {\cal M}_{RR} - 2 ({\cal M}^{(1)}_{R}+ {\cal M}^{(2)}_{R})
$.
This has the great advantage that one need only measure currents on one side of the dots. 
Furthermore, the electron counting we have pursued here only accounts for electrons leaving the dots and ignores those entering the dots in the initialisation phase. By expressing the measurement purely in terms of right-lead quantities, we avoid having to explicitly take the latter into account.
Figure \ref{FIG2}a shows the CHSH correlator, $F$,  as a function of time.  Violations of the CHSH inequality become visible after a time $\sim 4\Gamma^{-1}$

\begin{figure}[tb]
  \begin{center}
  \epsfig{file=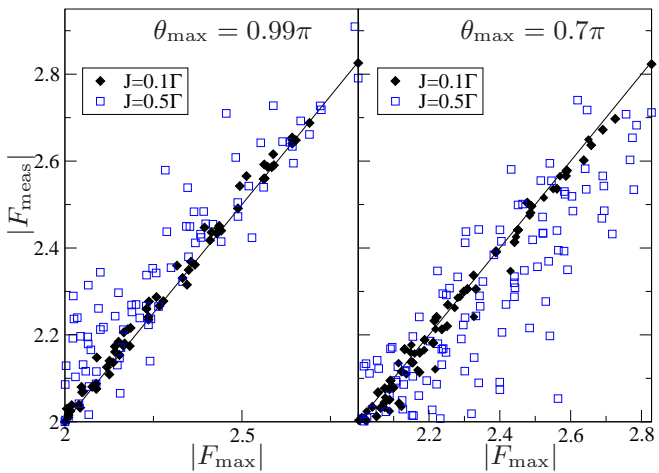, clip=true,width=\linewidth}
  \caption{
    As \fig{FIG2}b, but for randomly chosen density matrices.
    Ideal values of $|F_\mathrm{max}|$ calculated from Refs. \cite{hor95,ver02}.
    \label{FIG3}
  }
  \end{center}
\end{figure}

{\em Non-idealities.}
The situation in real dots differs from the preceding analysis in two respects that we now address: (i) the single-qubit Hamiltonians do not have a $\sigma_y$ component and are further restricted to experimentally accessible DQD parameters; and (ii) it is unlikely that the interaction between qubits can be completely suppressed during the measurement phase. 
The single-qubit part of $H_\mathrm{DQD}$ can be rewritten as
\beq
  H^{(i)} = \Lambda^{(i)} 
  \rb{
     \sin(\textstyle{\frac{1}{2}}\phi^{(i)}) \sigma_z^{(i)} 
     + 
     \cos(\textstyle{\frac{1}{2}}\phi^{(i)}) \sigma_x^{(i)}
  }
\eeq
with
$ \Lambda^{(i)}= \sqrt{\epsilon_i^2 + T_i^2} $.  If $T_i$ and $\epsilon_i$ are unrestricted in magnitude, angle $\phi^{(i)}$ has the range 
$-\pi \le \phi^{(i)} \le \pi$, and angle $\theta^{(i)}$ is bounded as
$0 \le \theta^{(i)} \le \pi$ with the upper bound attainable only in the 
$\Lambda^{(i)}/\Gamma \to \infty$ limit.  
With these bounds, a sufficient range of rotations can be performed to ensure maximum violation, despite the absence of the $\sigma_y$ component.  However, since $\Lambda^{(i)}/\Gamma$ is finite, the rotation angle is restricted, $0 \le \theta^{(i)} \le \theta^{(i)}_\mathrm{max}$, with  
$\theta_\mathrm{max}^{(i)} = 
  \textstyle{\frac{1}{2}} \arctan \rb{2 \Lambda^{(i)} / \Gamma}_\mathrm{max}$.  This constraint limits the optimisation of $F$.
More importantly, with interaction during the measurement, the value of $F$ obtained with the above scheme is no longer equal to the true value of $F$ of the state.
Figures \ref{FIG2} and \ref{FIG3} show the maximum value of $|F|$ that would be measured, $|F_\mathrm{meas}|$, against the value of $|F_\mathrm{max}|$ calculated from the state.
The measured results are obtained from numerical integration of the Master equation with the interacting Hamiltonian of \eq{DQDHam} and subsequent optimisation of $F$ subject to the above constraint on $\theta^{(i)}$.

Figures~\ref{FIG2}b and c shows the results for initial states belonging to the FLIP class.  In the noninteracting case ($J=0$) the restriction to  $\theta_\mathrm{max}^{(i)}<\pi$ causes only very small reductions in the measured value of $|F_\mathrm{max}|$ provided that $\theta_\mathrm{max}^{(i)}/\pi \gtrsim 0.7$, which corresponds to $  \rb{\Lambda^{(i)} / \Gamma}_\mathrm{max} \gtrsim 1$.
Furthermore, it is clear from the finite-$J$ results that, for FLIP-class states,  $|F_\mathrm{meas}|\le |F_\mathrm{max}|$ even with interaction present, and thus $|F_\mathrm{meas}|$, and the concurrence calculated from it, provide lower bounds for the actual values of $|F_\mathrm{max}|$ and ${\cal C}$
--- the measurement never flags an unentangled FLIP-class state as an entangled one.  Note further that the larger $\theta_\mathrm{max}$ is, the smaller the deviations. For example, at $\theta_\mathrm{max}/\pi=0.99$, even a value of $J=\Gamma$ produces only a deviation in $|F_\mathrm{max}|$ of $\sim 5\%$.
Figure \ref{FIG3} shows the same comparison 
for a set of randomly generated density matrices.  In this case, the measured value $|F_\mathrm{meas}|$ is no longer a lower bound for $|F_\mathrm{max}|$.  For small $J$, however, the measured values are strongly clustered around the ideal values, and tight upper and lower bounds can be derived.  For larger values of $J$, the clustering is not as tight, and unentangled states can give quite large values of $|F_\mathrm{meas}|$.  From these data, we infer that a reliable signature of entanglement is $|F_\mathrm{meas}| \gtrsim 2.35$ for $\theta_\mathrm{max}=0.99 \pi$ and $|F_\mathrm{meas}| \gtrsim 2.2$ for $\theta_\mathrm{max}=0.7 \pi$.  From this point of view, tighter restriction of  $\theta_\mathrm{max}$ may be advantageous.

From the foregoing, we can identify the parameter requirements for obtaining the strongest possible signal of entanglement.  In order to obtain tight bounds on  $F_\mathrm{max}$, we require $J< \Gamma \lesssim \Lambda_\mathrm{max}^{(i)}$ in the measurement phase.  Obtaining a high entanglement in the initialisation phase implies a different relation between $\Lambda^{(i)}$ and $J$ --- from \fig{FIG2} we observe that high concurrence occurs when $T_i \sim J$.  If we assume $\Lambda^{(i)}_\mathrm{max} \sim (T_i)_\mathrm{max} $, this implies that the single-qubit coupling should change from $T_i \sim J$ to $T_i > J$ between phases.  This may be feasible, but note that this result was obtained with a maximum initialisation time of $J^{-1}$. If we allow an initialisation time of $10 J^{-1}$, say, then due to the greater range of oscillation explored, high concurrences require $T_i \sim \Lambda^{(i)} \sim 0.1 J$, such that $T_i > J$ in both phases.  In the end, the maximum initialisation time is determined by the dephasing rate $\gamma$ and we require $\gamma \ll J$ for successful operation. For a dephasing time of 1ns \cite{hay03} (corresponding to $\gamma \sim 1 \mu$eV), and interaction strength $J=25\mu$eV, this relationship is satisfied.
This also implies that $\Gamma \gg \gamma$, consistent with the neglect of dehasing during the read-out phase.
Note that it might also be experimentally possible to increase the potential barrier  between the two DQDs at the end of the initialisation phase, thus separating the electrons and reducing their interaction.  This would further improve operation.

In summary, we have described a way of obtaining information on the entanglement of DQD charge qubits which uses shotnoise measurements of the decaying qubits to determine the CHSH parameter $F_\mathrm{max}$.  This BI approach does not need complete control over single-qubit rotations and it is not even necessary to know which rotations were performed. 
This can be contrasted with density matrix tomography \cite{Roo04}, which requires precise control and knowledge of the single-qubit rotations.
Our approach might also be used to measure DQD entanglement generated in other ways, e.g. through interaction with a common bath \cite{con08,stat}. 
Finally, we mention that the use of internal dynamics to generate BI rotations may have broader applicability to the study of entanglement in other decaying system, e.g. quantum optics.

\begin{acknowledgments}
This work was supported by the WE Heraeus foundation and by DFG grant BR 1528/5-1.  I am grateful to T.~Brandes for helpful discussions.
\end{acknowledgments}



\begin{thebibliography}{99}
\bibitem{stat}
    We considered the application of this scheme to an initial state given by the transport steady state of  Ref. \cite{lam07}.  We find that this state never violates the CHSH equality, even though its concurrence may be as high as  ${\cal C} = 0.3$.
\end{thebibliography}
\end{document}